\documentclass[preprint]{ptephy_v1}

\preprintnumber{OU-HET-868} 

\usepackage{graphics} 
\usepackage{color}
\usepackage{amsmath,amssymb}
\usepackage{bm}
\usepackage{ascmac}
\usepackage{def_simbol}

\newcommand{\bra}{\langle}
\newcommand{\ket}{\rangle}

\newcommand{\dsection}[1]{}
\renewcommand{\include}[1]{}

\begin{document}

\title{Effective lattice action for the configurations smeared by the Wilson flow}


\author{Aya Kagimura}
\author{Akio Tomiya}
\author{Ryo Yamamura}
\affil{Department of Physics, Osaka University, Toyonaka, Osaka 560-0043 Japan
\email{kagimura@het.phys.sci.osaka-u.ac.jp, akio@het.phys.sci.osaka-u.ac.jp, ryamamura@het.phys.sci.osaka-u.ac.jp}}


\begin{abstract}
We investigate a trajectory for the Wilson flow in the theory space.
For this purpose, we determine the coefficient of the plaquette and rectangular terms in the action for the configurations defined by the solution of the Wilson flow.
The demon method regarded as one of the inverse Monte Carlo methods
is used for the determination of them.
Starting from the conventional Wilson plaquette action of quenched QCD,
we find that the coefficient of the plaquette grows 
while that of the rectangular tends to negative with the development of the flow as the known improved actions.
We also find that the trajectory forms a straight line in the two-coupling theory space.
\end{abstract}

\subjectindex{B01, B32, B38}

\maketitle

\section{Introduction\label{introduction}}
The lattice gauge theory is one of the most powerful tool for investigating the quantum field theory. 
However, 
it is not suitable for studying quantities which are sensitive to
the cutoff: the topological charge, energy momentum tensor and so on.
The Wilson flow is considered to be a remedy for such issues
\cite{Luscher:2010iy, Luscher:2010we, Luscher:2011bx, Borsanyi:2012zs, Borsanyi:2012zr, Fodor:2012td, Fodor:2012qh, Fritzsch:2013je, Luscher:2013cpa}.
It is regarded as a continuous stout smearing and
one can measure such observables even on the discretized spacetime
\cite{Luscher:2009eq, Asakawa:2013laa}.
In Ref.~\cite{Luscher:2009eq}, M. L\"uscher has studied the effective action for the
configuration at finite $\hat{t}=t/a^2$ in the context of the transformation in the field space and given a analytic form of it.
The purpose of this paper is to determine it numerically by the so-called demon method.

The demon method has been proposed by M. Creutz~\cite{Creutz:1983ra} and
improved by M. Hasenbusch {\it et al.}~\cite{Hasenbusch:1994ne},
which enables us to determine effective couplings from a given configuration.
There has been various applications of this method.
For example,
investigating the couplings induced by the fermionic determinant~\cite{Blum:1994xb},
studying the renormalization group of the $SU(3)$ lattice gauge theory~\cite{Takaishi:1995ve}, and 
describing the $SU(3)$ lattice gauge theory in terms of effective Polyakov-loop models~\cite{Wozar:2007tz}.
Using the method, one can investigate effective actions without any perturbative approximation.

A similar work to ours has been done by QCD-TARO collaboration~\cite{deForcrand:1998xe, deForcrand:1999bi}.
They have studied the renormalization trajectory for 
 quenched QCD in the two-coupling theory space
performing the blocking of the link variables~\cite{Swendsen:1979gn} 
and using the Schwinger-Dyson equation method~\cite{GonzalezArroyo:1988dc}
to compute the effective coupling of the action for the blocked configurations.
As mentioned above, 
since the Wilson flow is regarded as a continuous smearing, 
we compare our effective action with the ones investigated in Ref.~\cite{deForcrand:1998xe, deForcrand:1999bi}.
As we will see later, our effective action shows the same tendency as the known improved action, 
however it has different pictures.

In this work, we determine a trajectory for the Wilson flow in the theory space using 
the demon method.
We choose $\beta=6.0$ Wilson plaquette action for configuration generation and apply the Wilson flow. 
The range of the flow time is taken to $0\leq \sqrt{8\hat{t}} \lesssim 1.3$.
In order to evaluate finite size effects, our lattice sizes are taken to $\hat{L}^4=(L/a)^4=4^4$, $8^4$ and $16^4$.
We find that the coefficient of the plaquette grows while that of the rectangular tends to negative with the flow time as the known improved actions. We also find that the trajectory forms a straight line in the two-coupling theory space.

This paper is organized as follows.
In Sec.~\ref{setup}, we briefly review the demon method 
and explain our strategy to obtain the effective action.
A numerical setup and results are shown in Sec.~\ref{numerical}.
Finally, a conclusion, discussions and future perspectives are given in Sec.~\ref{conclusion}.
\section{The demon method and our strategy\label{setup}}
\dsection{Setup}
In this section, we review the demon method \cite{Creutz:1983ra, Hasenbusch:1994ne}
and how we utilize it for the determination of the effective action for configurations
which are continuously smeared by the Wilson flow equation~\cite{Luscher:2010iy}:
\begin{align}
\label{eq:WilsonFloweq}
\dot{\bar{V}}_\mu(x;t)=
-g_0^2\{\partial_{x,\mu}S_{\rm W}(\bar{V})\}\bar{V}_\mu(x;t),~~\bar{V}_\mu(x;t)|_{t=0}=U_\mu(x),
\end{align}
where $S_{\rm W}(U)$ is defined by
\begin{align}
&~S_{\rm W}(U)=\beta\sum_{x,\mu<\nu}\bigg[1-\frac{1}{3}{\rm ReTr}~W^{1\times 1}_{\mu\nu}(x)\bigg],~~\beta=\frac{6}{g_0^2}, \notag \\
\label{eq:WilsonAction}
&~W^{1\times 1}_{\mu\nu}(x)=U_\mu(x)U_\nu(x+\mu)U^{\dag}_\mu(x+\nu)U^\dag_\nu(x).
\end{align}

For the definition of the effective action,
let us consider an expectation value of any operator $O[\bar{V}_t(U)]$
\begin{align}
\label{eq:flowed_observable}
\bra O \ket_t
= \frac{1}{Z}\int DU~O[\bar{V}_t(U)]~e^{-S_{\rm W}[U]},
\end{align}
where 
\begin{align}
\label{eq:}
Z=\int DU~e^{-S_{\rm W}[U]},
\end{align}
and $\bar{V}_t\equiv \{\bar{V}_\mu(x;t)\}$.
The effective action is defined by inserting the delta function and 
changing the integration variables from $U$ to $V_t$ as follows.
\begin{align}
\label{eq:observable_eff}
\bra O\ket _t
&~=\frac{1}{Z}\int DU
\bigg[ \int DV_t ~\delta(V_t-\bar{V}_t(U)) \bigg]
~O[\bar{V}_t(U)]~e^{-S_{\rm W}[U]} \notag \\
&~=\frac{1}{Z}\int DUDV_t ~\delta(V_t-\bar{V}_t(U)) 
~O[V_t]~e^{-S_{\rm W}[U]} \notag \\
&~\equiv\frac{1}{Z_{\rm eff}}\int DV_t  
~O[V_t]~e^{-S_{\rm eff}[V_t]}, 
\end{align}
where 
\begin{align}
Z_{\rm eff}=\int DV_t~e^{-S_{\rm eff}[V_t]}.
\end{align}
$S_{\rm eff}$ can be regarded as the ``improved" action
since the discretization effects for observables are reduced at finite $\hat t$. 

\subsection{The demon method\label{demon}}
\dsection{The demon method}
Here, we briefly review 
the demon method which enables us to determine the couplings of the action from a given configuration.

Let us consider a configuration whose distribution is the Boltzmann weight
\begin{align}
\label{eq:boltzmann}
P(U) \propto e^{-\beta S[U]},
\end{align}
where $\beta$ is {\it a priori} unknown coupling 
and to be determined by the demon method.

Now we introduce an extra degree of freedom ``demon" to the system
and consider a microcanonical partition function of a joint system,
\begin{align}
Z_{\rm mic} = \sum_U\sum_{E_d}\delta(S[U]+E_d-E_0),
\end{align}
where $E_0$ is an initially determined total energy of the joint system,
and $E_d$ is the energy carried by the demon.
The demon energy is restricted within the range $E_{\rm min}<E_d<E_{\rm max}$ which can be chosen suitably.
Hereafter we set $E_{\rm min} = -E_{\rm max}$ for simplicity. 
Starting from a given configuration,
we update the system keeping $S[U] + E_d$ constant.
The updating process is described as follows.
First, a new configuration $U'$ is proposed.
To keep the total energy constant,
the new demon energy is given by
\begin{align}
E_d'=E_d-S[U']+S[U].
\end{align}
When $E_d'$ is in the allowed region [$-E_{\rm max}$:$E_{\rm max}$],
the new configuration and the new demon energy are accepted,
otherwise they are rejected and the configuration and the demon energy remain to be same.
If the probability of the change is symmetric in $\{U,E_d\}$ and $\{U',E_d'\}$,
a generated sequence of combined configurations is distributed according to a uniform distribution.

When the degrees of freedom of $U$ are sufficiently large,
a given configuration behaves as a heat bath to thermalize the demon.
Therefore standard statistical mechanics arguments show that $E_d$ is distributed with
\begin{align}
P(E_d) \propto e^{-\beta E_d},
\end{align}
in the thermodynamic limit.
The average of $E_d$ is given by
\begin{align}
\label{eq:Ed}
\bra E_d\ket
=\frac{1}{Z}\int_{-E_{\text{max}}}^{E_{\text{max}}}dE_d~E_d~e^{-\beta E_d}
=\frac{1}{\beta}
\bigg[1-\frac{\beta E_{\text{max}}}{\tanh(\beta E_\text{max})}\bigg],
\end{align}
where 
\begin{align}
Z = \int_{-E_{\text{max}}}^{E_{\text{max}}}dE_d~e^{-\beta E_d}.
\end{align}
Finally, $\beta$ is obtained by measuring $\bra E_d\ket$ and solving Eq.~\eqref{eq:Ed}.

The extension to a multi-coupling system is straightforward.
Suppose that the action is parametrized by
\begin{align}
\label{eq:multi}
S[U]=\sum_i\beta_iS_i[U],
\end{align}
where $S_i[U]$ is the interaction term and $\beta_i$ is the corresponding coupling.
Now we introduce demons $E_d^i$ for each coupling.
During the microcanonical update, a proposed configuration and demon energies are accepted
only if all demon energies are in the allowed region.
Here we take the common allowed region $[-E_{\rm max}:E_{\rm max}]$ to all couplings for simplicity.
$\beta_i$ is obtained by solving
\begin{align}
\label{eq:beta_multi}
\bra E_d^i\ket
=\frac{1}{\beta_i}\bigg[1-\frac{\beta_i E_{\text{max}}}{\tanh(\beta_i E_\text{max})}\bigg].
\end{align}
\subsection{Our strategy\label{strategy}}
\begin{figure}[htbp]
\centering
\includegraphics[width=12cm,bb=0 0 650 388]{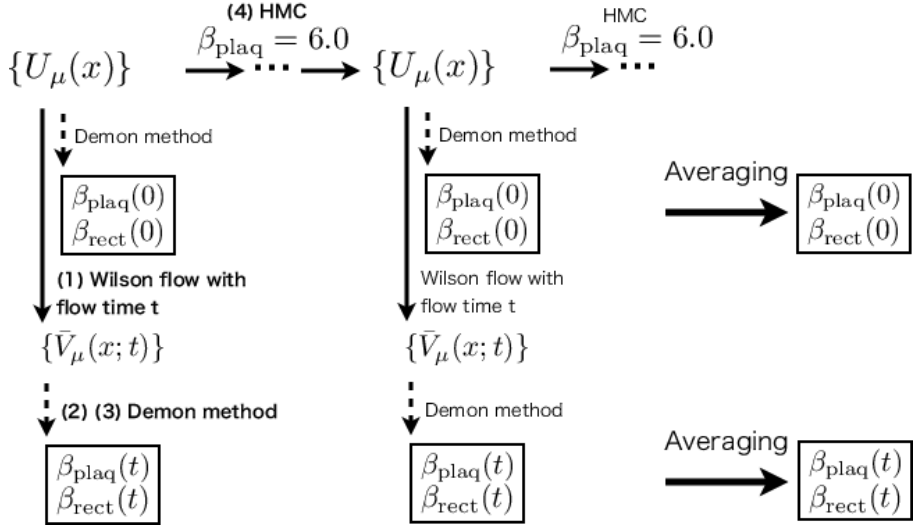}
\caption{Schematic picture of our strategy.
Bold texts with the number correspond to the explanation in the text.}
\label{fig:strategy_schematic}
\end{figure}

\dsection{Strategy to get the effective action for Wilson flowed configurations}
In this subsection, 
we explain our strategy to obtain the effective action defined by Eq.~\eqref{eq:observable_eff}.
We assume that $S_{\rm eff}$ 
can be parametrized by the form of Eq.~(\ref{eq:multi}), 
where $S_i[U]$ denotes several types of Wilson loops such as the plaquette, rectangular, chair, sofa, etc.
The simulation is implemented as the following steps.
\begin{enumerate}
\item Evolve a given configuration by integrating the Wilson flow equation Eq.~\eqref{eq:WilsonFloweq} from 0 to $\hat t$.
\item Perform the microcanonical updates for the joint system using the configuration at $\hat t$ according to the procedure described in Subsec.~\ref{demon}.
\item After a sufficiently large number of microcanonical updates, 
take an average of the demon energy and obtain $\beta_i$ by solving Eq.~\eqref{eq:beta_multi}.
\item Replace the configuration at $\hat t=0$ by a new statistically independent one.
This is done by the HMC updates of the original system at $\hat t=0$.
In this step the demon degrees of freedom are frozen.
\item Go back to the step (1) with the new configuration.
The initial demons' energies in the step (2) are given by the averages of the previous run.
\end{enumerate}
The flowchart of the above algorithm is shown schematically in Fig.~\ref{fig:strategy_schematic}.
By taking averages of $\beta_i$s obtained for each gauge configuration,
we finally determine the couplings of the effective action.
The step (4) and (5) are needed to suppress the systematic errors coming from the finiteness of the volume
\cite{Hasenbusch:1994ne}.

\section{Numerical setup and results\label{numerical}}
\subsection{Numerical setup}
\dsection{numerical setup,tomiya}
Here, we introduce our numerical setup.
We employ the Wilson plaquette action defined by Eq.~\eqref{eq:WilsonAction} with $\beta=6.0$ for configuration generations.
Configurations are generated by the HMC algorithm.
Our lattice sizes are $(L/a)^4$=$\hat{L}^4=4^4$, $8^4$, and $16^4$.

In this work, we implement the demon method with the truncation
\begin{align}
S_{\rm eff}(U)=\bplaq \sum_{x,\mu<\nu}
\bigg[1-\frac{1}{3}{\rm ReTr~}W^{1\times 1}_{\mu\nu}(x)\bigg]
+\brect \sum_{x,\mu\neq\nu}\bigg[1-\frac{1}{3}{\rm ReTr~}R^{1\times 2}_{\mu\nu}(x)\bigg],
\end{align}
where $R^{1\times 2}_{\mu\nu}(x)$ is the rectangular loop defined by
\begin{align}
\label{eq:Rect}
R^{1\times 2}_{\mu\nu}(x)=U_\mu(x)U_\mu(x+\mu)U_\nu(x+2\mu)U^\dag_\mu(x+\nu+\mu)U^\dag_\mu(x+\nu)U^\dag_\nu(x).
\end{align}

Each gauge configuration is separated by 10 trajectories of the HMC.
The total number of measurements is 500, 500 and 250
for $\hat L^4=4^4$, $8^4$ and $16^4$ respectively.
The autocorrelation times defined by the average of the plaquette of each configuration 
are around 4.6, 11 and 11 for $\hat L^4=4^4$, $8^4$ and $16^4$.

Our measurements are performed as follows. 
All of the initial demon energies are set to be $0$ and $E_{\rm max}$ is set to be 5.0.
We perform 10000 microcanonical updates of the joint system 
with a given trial configuration for the thermalization of the demon energies. 
First 2000 updates are discarded and 
the averages of the demon energies over the last 8000 updates are taken.
These averages are used as the initial demon energies for the next microcanonical updates. 
Following steps are repeated during the measurements.
After a replacement of the gauge configuration, 
we always perform the 10000, 5000 and 5000 microcanonical updates for $\hat L^4 = 4^4$, $8^4$ and $16^4$.
First 2000, 1000 and 1000 updates are discarded and 
we take the averages of the demon energies over the last 8000, 4000 and 4000 updates 
for $\hat L^4=4^4$, $8^4$ and $16^4$ respectively.
The above procedures are implemented for each flow time.
The range of the flow time is taken to $0\leq \sqrt{8\hat{t}} \lesssim 1.3$.
\subsection{Results\label{results}}
We show the results of the flow time dependence of the effective action in this subsection.
The values of $\beta_{\rm plaq}$ and $\beta_{\rm rect}$
for $0\leq\hat{t}\leq0.20$ are listed in Table.~\ref{tab:results}.
We examine the consistency between the effective action for $\hat t = 0$ and the initial action.
Since our initial couplings $(\beta_{\rm plaq},\beta_{\rm rect}) = (6.0,0.0)$ are reconstructed 
within 2 sigma, the demon method works well.

\begin{figure}[tbhp]
\centering
\includegraphics[width=12cm]{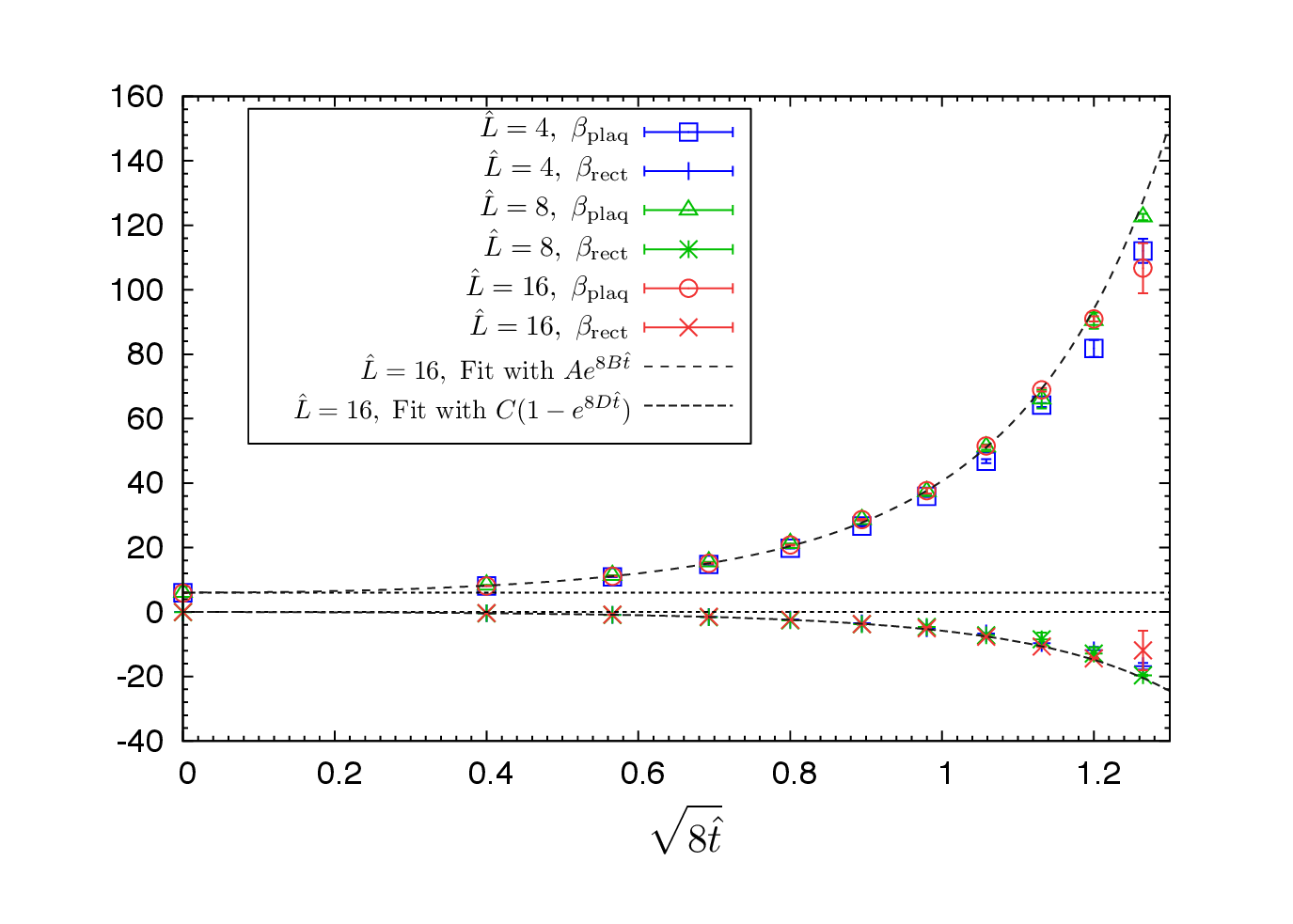}
\vspace{-1em}
\caption{The values of $\beta_{\rm plaq}$ and $\beta_{\rm rect}$
as a function of $\sqrt{8\hat{t}}$.
Two dashed horizontal lines are at the initial value of $\bplaq=6.0$ and $\brect=0$ respectively.
Two curved lines are the fit lines for $\hat{L}^4=16^4$.
The details of the fit are described in the text.}
\label{fig:couplings}
\end{figure}

The results show that the value of $\beta_{\rm plaq}$ grows with $\hat{t}$ (Fig. \ref{fig:couplings}).
We expect that this is because the Wilson flow has a smoothing effect on the gauge field and 
lowers the value of the Wilson action as $\hat{t}$ increases~\cite{Luscher:2009eq}.

On the other hand, the value of $\beta_{\rm rect}$ tends to a negative region and decreases with $\hat{t}$ 
(Fig. \ref{fig:couplings}).
This fact seems reasonable since 
the negativeness of $\brect$ is a common feature of the known improved actions 
such as the Symanzik~\cite{Weisz:1982zw, Weisz:1983bn, Luscher:1984xn}, 
Iwasaki~\cite{Iwasaki:1983ck, Iwasaki:1985we} and DBW2 action~\cite{deForcrand:1999bi}.
Note that our scheme is not based on any perturbative analysis 
and therefore defines the non-perturbatively ``improved" action. 
Moreover, 
the action may be systematically improved 
by adding conceivable Wilson loops to the ansatz of $S_{\rm eff}$.

\begin{table}[htb]
\begin{center}
\begin{tabular}{c|cc|cc|cc} \hline
& \multicolumn{2}{|c|}{$\hat L^4=4^4$} & \multicolumn{2}{|c|}{$\hat L^4=8^4$} & \multicolumn{2}{|c}{$\hat L^4=16^4$} \\ \hline
$\hat{t}$& $\beta_{\rm plaq}$ & $\beta_{\rm rect}$ & $\beta_{\rm plaq}$ & $\beta_{\rm rect}$ & $\beta_{\rm plaq}$ & $\beta_{\rm rect}$ \\ \hline
0.00 & 5.972(23) & 0.009(9) & 6.072(38) & -0.024(13) & 5.932(66) & 0.030(24) \\ \hline
0.02 & 8.116(44) & -0.363(12) & 8.368(75) & -0.387(17) & 8.087(58) & -0.329(25) \\ \hline
0.04 & 10.856(59) & -0.837(17) & 11.334(49) & -0.895(20) & 11.088(70) & -0.817(30) \\ \hline
0.06 & 14.741(83) & -1.490(19) & 15.512(63) & -1.602(24) & 15.120(141) & -1.440(69) \\ \hline
0.08 & 19.789(86) & -2.329(21) & 21.204(82) & -2.565(30) & 20.749(148) & -2.408(41) \\ \hline
0.10 & 26.671(189) & -3.477(39) & 28.637(200) & -3.791(72) & 28.660(243) & -3.705(56) \\ \hline
0.12 & 35.881(190) & -5.002(34) & 37.32(117) & -4.622(701) & 37.622(841) & -5.045(347) \\ \hline
0.14 & 46.793(619) & -6.682(152) & 51.208(848) & -7.207(523) & 51.535(289) & -7.692(75) \\ \hline
0.16 & 64.23(600) & -9.635(127) & 66.03(287) & -8.48(205) & 68.986(557) & -10.711(132) \\ \hline
0.18 & 81.79(260) & -11.88(101) & 90.41(251) & -12.86(185) & 90.930(801) & -14.408(207) \\ \hline
0.20 & 112.06(373) & -16.83(101) & 122.56(99) & -19.663(216) & 106.69(776) & -11.92(613) \\ \hline
\end{tabular}
\caption{The values of $\beta_{\rm plaq}$ and $\beta_{\rm rect}$ at $0\leq \hat t \leq 0.20$ for $\hat{L^4}=4^4,8^4$ and $16^4$.}
\label{tab:results}
\end{center}
\end{table}

$\bplaq$ and $\brect$ are plotted in Fig.~\ref{fig:couplings}, 
where the horizontal axis indicates $\sqrt{8\hat{t}}$, which is the effective range of smearing the link variables. 
Since the values of $\bplaq$ and $\brect$ for each $\hat{L}$ at a fixed $\hat{t}$ coincide with each other,
finite volume effects turn out to be irrelevant.

$\beta_{\rm plaq}$ and $\beta_{\rm rect}$ for $\hat L^4=16^4$ are fitted in the forms
\begin{align}
\label{eq:fit}
\beta_{\rm plaq}(\hat t)=Ae^{8B\hat t},~\beta_{\rm rect}(\hat t)=C(1-e^{8D\hat t}),
\end{align}
yielding the numerical results 
\begin{align}
\label{eq:fitresult}
A=6.040(47),~B=1.906(7),~C=1.010(47),~D=1.911(35),
\end{align}
and plotted with the curved lines respectively in Fig.~\ref{fig:couplings}.
This result shows that 
the $\hat t$ dependence of the exponent is common to these two couplings within the errors.

\begin{figure}[htbp]
\begin{center}
\includegraphics[width=12cm]{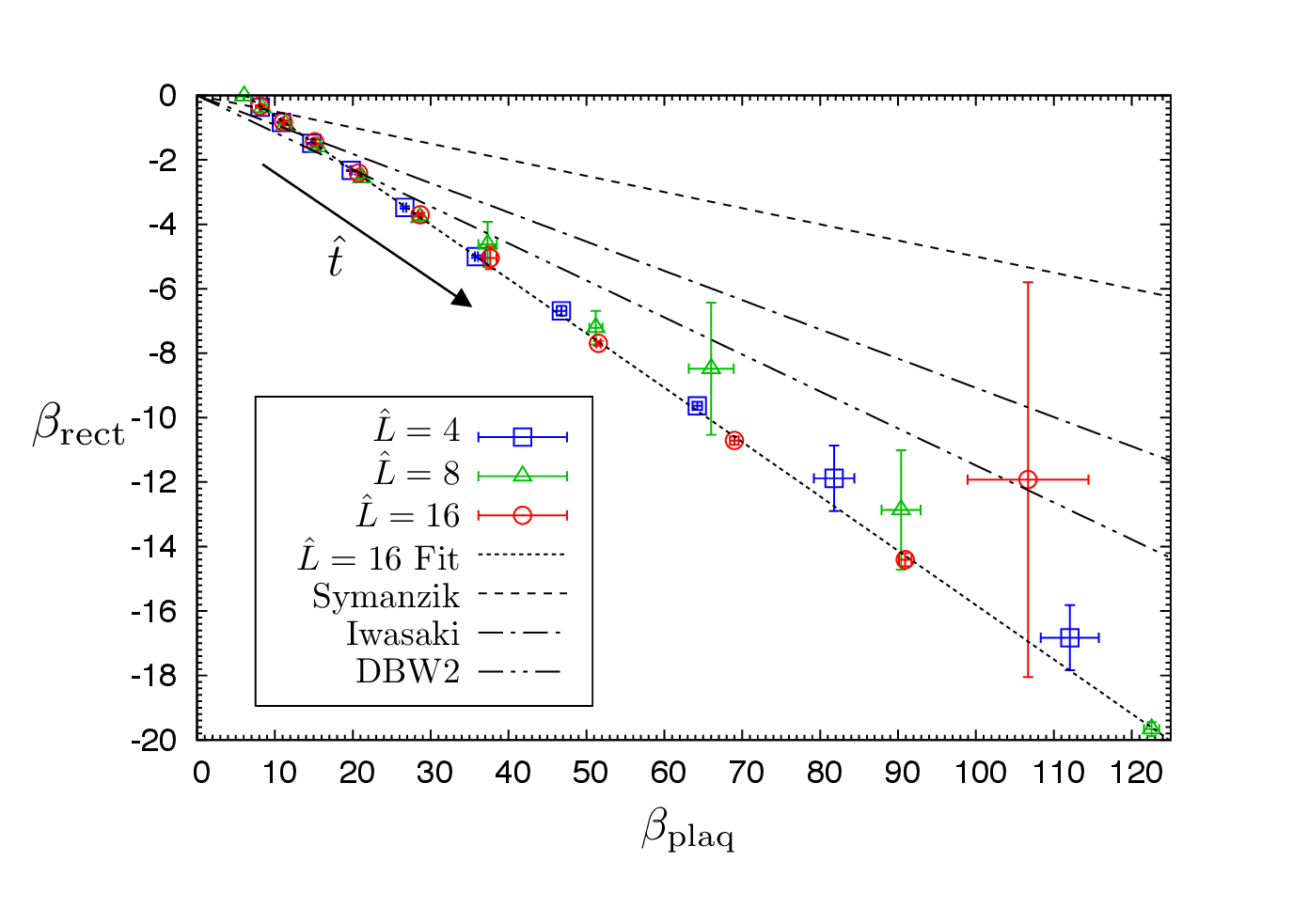}
\vspace{-1em}
\caption{The flow of the effective action in two-coupling theory space.
The value of $\hat t$ becomes larger along the arrow.
The doted line represents our numerical fit for $\hat L^4=16^4$. 
Other lines represent
the Symanzik~\cite{Weisz:1982zw, Weisz:1983bn, Luscher:1984xn}, 
Iwasaki~\cite{Iwasaki:1983ck, Iwasaki:1985we} and DBW2 action~\cite{deForcrand:1999bi}.
We take $-\brect/\bplaq=0.05,~0.09073~{\rm and}~0.11481$ for 
the Symanzik, Iwasaki and DBW2 action respectively.
}
\label{fig:cplot}
\end{center}
\end{figure}

\begin{figure}[htbp]
\begin{center}
\includegraphics[width=12cm]{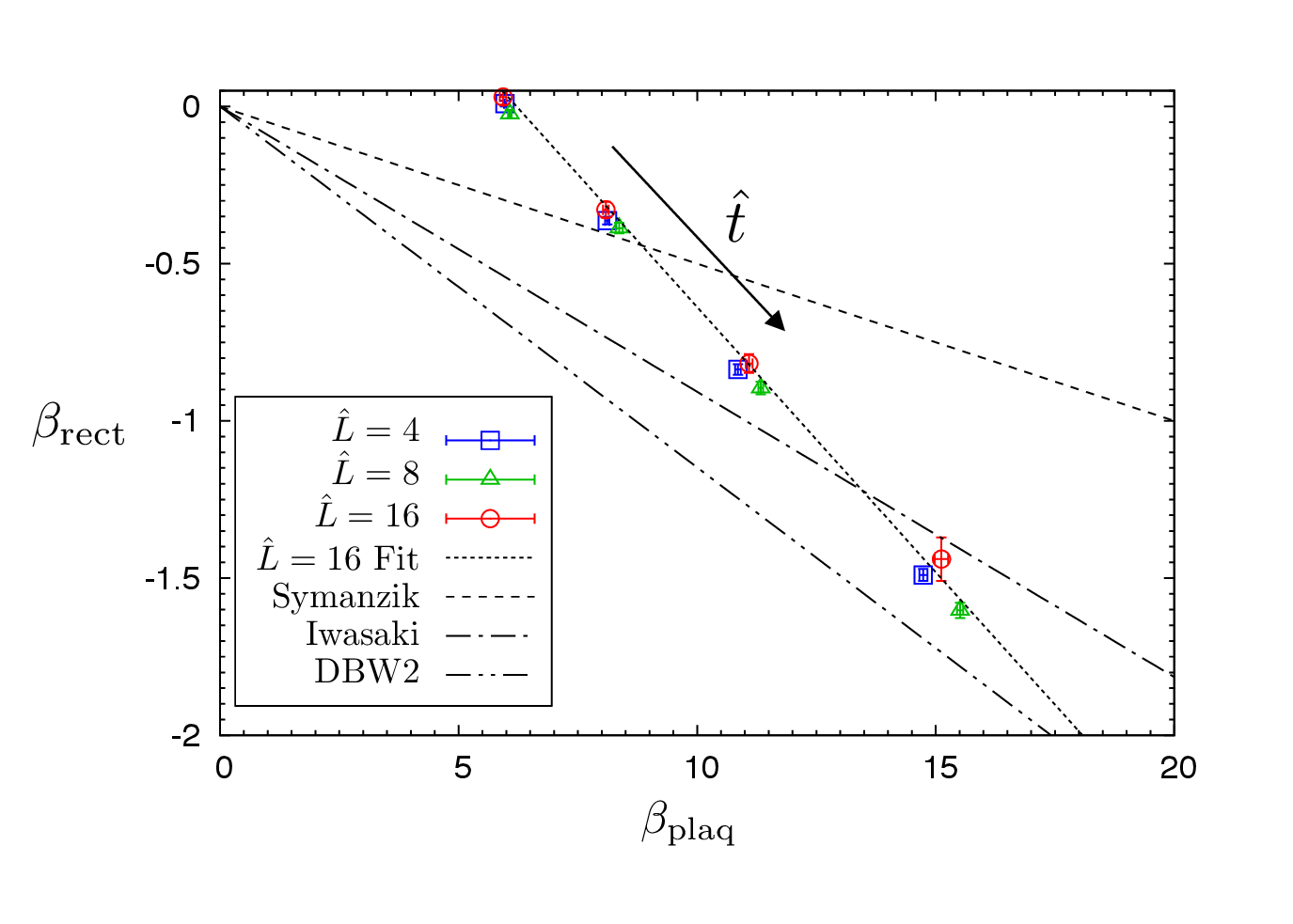}
\vspace{-1em}
\caption{The same figure as Fig.~\ref{fig:cplot} but around the origin.}
\label{fig:cplot_kakudai}
\end{center}
\end{figure}

We also plot the flow of the effective action in the two-coupling theory space in Fig.~\ref{fig:cplot}.
Fig.~\ref{fig:cplot_kakudai} shows the plot around the origin of Fig.~\ref{fig:cplot}.
Reflecting the fact that two couplings have the common exponent as the function of $\hat t$,
the flow of the effective action forms a non-trivial straight line (Fig.~\ref{fig:cplot}):
\begin{align}
\label{eq:eqline}
\bplaq+5.931(31)\brect=6.216(121),
\end{align}
determined by a numerical fitting for $\hat L^4=16^4$ with $\chi^2/{\rm d.o.f.} = 0.857(371)$.



\section{Conclusion and Discussions\label{conclusion}}
\dsection{Conclusion and discussions}
We have investigated a trajectory for the Wilson flow in the theory space by the demon method.
The effective action is truncated such that it has the plaquette and rectangular terms.
We have measured the flow time dependence of $\bplaq$ and $\brect$,
and found that $\bplaq$ increases while $\brect$ tends to the negative region with $\hat{t}$.
As shown in Fig.~\ref{fig:cplot} and Fig.~\ref{fig:cplot_kakudai},
we have found that the trajectory forms the straight line in two-coupling theory space,
although we do not know whether this is just an accident or not.
Since the value of $\beta$ has been fixed throughout this work,
it should be examined whether this fact is universal to all values of $\beta$.
A trajectory in a wider parameter space has to be investigated
by adding more Wilson loops to the ansatz as a next step of this work.
It is also interesting to study how the trajectory is changed when we employ other actions for the gradient flow.

As noted in Sec.~\ref{introduction},
one can measure the quantities which are sensitive to the cutoff by using the Wilson flow. 
From this fact, our effective action is expected to be an ``improved" action 
which reduces the discretization effects.
Actually, we have found that the flow time dependence of $\bplaq$ and $\brect$ shows 
the same tendency as in the already known improved actions
~\cite{Weisz:1982zw, Weisz:1983bn, Luscher:1984xn, Iwasaki:1983ck, Iwasaki:1985we, 
deForcrand:1999bi}. 
In order to confirm this, 
the cutoff sensitive quantities, such as the topological charge and the energy momentum tensor, 
should be studied using our effective action.
It is an interesting question whether such quantities are well-measured by tuning only two parameters in the action.
The restoration of the rotational symmetry should also be checked 
by measuring, for example, the difference between the on-axis and off-axis Wilson loops
\cite{deForcrand:1998xe, deForcrand:1999bi}.
The advantages of our effective action are as follows.
First, $S_{\rm eff}$ itself is totally well-defined without any ambiguity
such as truncations in blocking and projections of the link variables.
Second, it does not require any perturbative analysis.
Finally, we have one tunable parameter: the flow time.


Related to the improvement of the lattice action,
lattice artifacts could be reduced by the Wilson flow due to its smearing property.
There is a well-known spurious UV fixed point caused by lattice artifacts
when we have the adjoint plaquette term or the fermionic determinant~\cite{Blum:1994xb}.
The demon method combined with the Wilson flow, 
which is explained in Subsec.~\ref{strategy} can also be applied to this issue.
Namely, one can check the issue by measuring the coupling of the adjoint plaquette term of configurations
 generated by the fundamental-adjoint mixed action.

Since the Wilson flow is similar to a continuous ``block spin transformation",
it may be used to define the exact renormalization group of the lattice gauge theories.
However, the trajectory for the Wilson flow in the theory space 
travels in the opposite direction to the renormalization trajectory 
investigated by QCD-TARO collaboration~\cite{deForcrand:1998xe, deForcrand:1999bi}. 
Therefore, our effective action
cannot be regarded as an action obtained after a renormalization group transformation.
This is because the Wilson flow itself is just a ``blocking" and 
does not contain a rescaling, which is necessary for the 
renormalization group transformation.
Just replacing the blocking procedure~\cite{Swendsen:1979gn}
used in Ref.~\cite{deForcrand:1998xe, deForcrand:1999bi}
by the Wilson flow is considered as a way to construct a renormalization group scheme via the Wilson flow.
This is also an interesting subject related to our work.

Finally, we comment on two types of systematic errors in this work.

One stems from the truncation of the effective action. 
This can be reduced by adding possible Wilson loops to our truncated action.
Since the configurations at large $\hat t$ picks up the information of the original configurations at widely separated points,
larger Wilson loops should be required with increasing the flow time.
Therefore our truncation may be reasonable only in the small $\hat{t}$ regime. 

Another comes from the demon method itself.
As noted in Subsec.~\ref{results}, the Wilson flow lowers the value of the Wilson action as $\hat t$ increases.
Since the gauge configurations are changed randomly during the microcanonical updates,
most of attempts raise the value of $S_{\rm W}$ and are rejected.
Therefore it is hard to determine a large-valued $\beta_i$ precisely in the demon method.
This error is unavoidable as long as using the demon method, 
however there is another way to determine the couplings from a given configuration using 
the Schwinger-Dyson equation method~\cite{GonzalezArroyo:1988dc}.

\ack
We thank Hidenori Fukaya, Hideo Matsufuru and Tetsuya Onogi for fruitful discussions and useful comments. 
We also thank Hiroshi Suzuki for his warm encouragements.
Guido Cossu, Hideo Matsufuru, Jun-ichi Noaki and Satoru Ueda 
gave us  generous support for the setup of the numerical calculations.
This work is in part based on Bridge++ code~\cite{Ueda:2014zsa} (http://bridge.kek.jp/Lattice-code/) and Iroiro++ code
~\cite{Cossu:2013ola}.
The numerical calculations were carried out on SR16000 at YITP in Kyoto University.
R.Y. is supported in part by the Grand-in-Aid of the Japanese Ministry of Education (No. 15J01081).

\bibliographystyle{ptephy}
\bibliography{bib}

\begin{thebibliography}{10}

\bibitem{Luscher:2010iy}
Martin Luscher, JHEP, {\bf 08}, 071, [Erratum: JHEP03,092(2014)] (2010),
  {{arXiv:1006.4518}}.

\bibitem{Luscher:2010we}
Martin Luscher, PoS, {\bf LATTICE2010}, 015 (2010),  {{arXiv:1009.5877}}.

\bibitem{Luscher:2011bx}
Martin Luscher and Peter Weisz, JHEP, {\bf 02}, 051 (2011),
  {{arXiv:1101.0963}}.

\bibitem{Borsanyi:2012zs}
Szabolcs Borsanyi et~al., JHEP, {\bf 09}, 010 (2012),  {{arXiv:1203.4469}}.

\bibitem{Borsanyi:2012zr}
Szabolcs Borsanyi, Stephan Durr, Zoltan Fodor, Sandor~D. Katz, Stefan Krieg,
  Thorsten Kurth, Simon Mages, Andreas Schafer, and Kalman~K. Szabo (2012),
  {{arXiv:1205.0781}}.

\bibitem{Fodor:2012td}
Zoltan Fodor, Kieran Holland, Julius Kuti, Daniel Nogradi, and Chik~Him Wong,
  JHEP, {\bf 11}, 007 (2012),  {{arXiv:1208.1051}}.

\bibitem{Fodor:2012qh}
Zoltan Fodor, Kieran Holland, Julius Kuti, Daniel Nogradi, and Chik~Him Wong,
  PoS, {\bf LATTICE2012}, 050 (2012),  {{arXiv:1211.3247}}.

\bibitem{Fritzsch:2013je}
Patrick Fritzsch and Alberto Ramos, JHEP, {\bf 1310}, 008 (2013),
  {{arXiv:1301.4388}}.

\bibitem{Luscher:2013cpa}
Martin Luscher, JHEP, {\bf 1304}, 123 (2013),  {{arXiv:1302.5246}}.

\bibitem{Luscher:2009eq}
Martin Luscher, Commun.Math.Phys., {\bf 293}, 899--919 (2010),
  {{arXiv:0907.5491}}.

\bibitem{Asakawa:2013laa}
Masayuki Asakawa, Tetsuo Hatsuda, Etsuko Itou, Masakiyo Kitazawa, and Hiroshi
  Suzuki, Phys. Rev., {\bf D90}(1), 011501 (2014),  {{arXiv:1312.7492}}.

\bibitem{Creutz:1983ra}
Michael Creutz, Phys.Rev.Lett., {\bf 50}, 1411 (1983).

\bibitem{Hasenbusch:1994ne}
M.~Hasenbusch, K.~Pinn, and C.~Wieczerkowski, Phys.Lett., {\bf B338}, 308--312
  (1994),  {{arXiv:hep-lat/9406019}}.

\bibitem{Blum:1994xb}
T.~Blum, Carleton~E. DeTar, Urs~M. Heller, Leo Karkkainen, K.~Rummukainen, and
  D.~Toussaint, Nucl. Phys., {\bf B442}, 301--316 (1995),
  {{arXiv:hep-lat/9412038}}.

\bibitem{Takaishi:1995ve}
T.~Takaishi, Mod. Phys. Lett., {\bf A10}, 503--514 (1995).

\bibitem{Wozar:2007tz}
Christian Wozar, Tobias Kaestner, Andreas Wipf, and Thomas Heinzl, Phys. Rev.,
  {\bf D76}, 085004 (2007),  {{arXiv:0704.2570}}.

\bibitem{deForcrand:1998xe}
P.~de~Forcrand et~al. (1998),  {{arXiv:hep-lat/9806008}}.

\bibitem{deForcrand:1999bi}
P.~de~Forcrand, M.~Garcia~Perez, T.~Hashimoto, S.~Hioki, H.~Matsufuru,
  O.~Miyamura, A.~Nakamura, I.~O. Stamatescu, T.~Takaishi, and T.~Umeda, Nucl.
  Phys., {\bf B577}, 263--278 (2000),  {{arXiv:hep-lat/9911033}}.

\bibitem{Swendsen:1979gn}
R.~H. Swendsen, Phys. Rev. Lett., {\bf 42}, 859--861 (1979).

\bibitem{GonzalezArroyo:1988dc}
Antonio Gonzalez-Arroyo and J.~Salas, Phys. Lett., {\bf B214}, 418 (1988).

\bibitem{Weisz:1982zw}
P.~Weisz, Nucl. Phys., {\bf B212}, 1 (1983).

\bibitem{Weisz:1983bn}
P.~Weisz and R.~Wohlert, Nucl. Phys., {\bf B236}, 397, [Erratum: Nucl.
  Phys.B247,544(1984)] (1984).

\bibitem{Luscher:1984xn}
M.~Luscher and P.~Weisz, Commun. Math. Phys., {\bf 97}, 59, [Erratum: Commun.
  Math. Phys.98,433(1985)] (1985).

\bibitem{Iwasaki:1983ck}
Y.~Iwasaki (1983).

\bibitem{Iwasaki:1985we}
Y.~Iwasaki, Nucl. Phys., {\bf B258}, 141--156 (1985).

\bibitem{Ueda:2014zsa}
S.~Ueda, S.~Aoki, T.~Aoyama, K.~Kanaya, H.~Matsufuru, S.~Motoki, Y.~Namekawa,
  H.~Nemura, Y.~Taniguchi, and N.~Ukita, PoS, {\bf LATTICE2013}, 412 (2014).

\bibitem{Cossu:2013ola}
Guido Cossu, Jun Noaki, Shoji Hashimoto, Takashi Kaneko, Hidenori Fukaya,
  Peter~A. Boyle, and Jun Doi,
\newblock {JLQCD IroIro++ lattice code on BG/Q},
\newblock In {\em {Proceedings, 31st International Symposium on Lattice Field
  Theory (Lattice 2013)}} (2013),  {{arXiv:1311.0084}}.

\end{thebibliography}

\end{document}